\documentclass{jfm}
\usepackage{epstopdf, epsfig}
\usepackage{siunitx}
\usepackage{url}
\usepackage{xcolor}
\usepackage{bm}
\usepackage{newtxtext}
\usepackage[varg]{newtxmath}

\shorttitle{Microstructure and thickening of dense suspensions under extensional and shear flows}
\shortauthor{R. Seto, G. G. Giusteri, A. Martiniello}
\title{Microstructure and thickening of dense suspensions under extensional and shear flows}
\author{Ryohei Seto\aff{1}
  \corresp{\email{setoryohei@me.com}},
  Giulio G. Giusteri\aff{1},
  Antonio Martiniello\aff{1} 
}

\affiliation{\aff{1}Mathematical Soft Matter Unit, Okinawa Institute of Science and Technology Graduate University, 1919-1 Tancha, Onna, Okinawa, 904-0495, Japan}

\DeclareMathOperator{\tr}{Tr}
\newcommand*\Tensor[1]{\mathsfbi{#1}}
\newcommand{\cddot}{\mathbin{:}}
\newcommand{\secref}[1]{\S\ref{#1}}

\newcommand{\figref}[1]{\figurename~\ref{#1}}

\newcommand\StNum{\mbox{St}}  
\newcommand{\CauchyStress}{\bm{\sigma}}

\begin{document}
\maketitle

\begin{abstract}
Dense suspensions are non-Newtonian fluids 
which exhibit strong shear thickening and normal stress differences.
Using numerical simulation of extensional and shear flows,
we investigate how rheological properties are determined by the microstructure 
which is built under flows and by the interactions between particles.
By imposing extensional and shear flows,
we can assess the degree of flow-type dependence in
regimes below and above thickening.
Even when the flow-type dependence is hindered,
non-dissipative responses, such as normal stress differences, 
are present and characterise the non-Newtonian behaviour of dense suspensions.

\end{abstract}


\section{Introduction}

Suspensions, namely mixtures of solid particles and a viscous liquid,
can be considered as an incompressible fluid 
as long as the volume fraction $\phi$ of solid particles
is less than a certain value, the jamming point, above which a solid-like behaviour is observed.  
The behaviour of suspensions is not usually captured by simple Newtonian models.
As primary example of non-Newtonian effect, 
the viscosity can vary with the shear rate, exhibiting shear thinning and shear thickening~\citep{Laun_1984,Barnes_1989,Bender_1996,Guy_2015}.
Moreover, nonvanishing normal stress differences $N_1$ and $N_2$, another hallmark of non-Newtonian behaviour,
are often observed~\citep{Laun_1994,Lootens_2005,Lee_2006a,Couturier_2011,Dbouk_2013,Cwalina_2014}.
Discontinuous shear thickening is a particularly intriguing phenomenon of dense suspensions and the underlying mechanism raised a significant debate~\citep{Brady_1985,Hoffmann_1998,Melrose_2004,Fall_2008,Brown_2009}.
Analysing the 
rheology of suspensions is a difficult task since
forces of various nature act among particles and the system lives mostly far from thermodynamic equilibrium.
Particle simulations have been used to explore the microstructure emerging among particles in various flows
and to estimate the importance of different interactions.
Several particle simulations recently succeeded
in reproducing shear thickening
by taking into account direct contact forces%
~\citep{Fernandez_2013,Heussinger_2013,Seto_2013a}.
These works support the ``stress-induced friction'' scenario~\citep{Wyart_2014,Mari_2014} and 
the contribution of contact forces
was also confirmed in experiments~\citep{Lin_2015,Clavaud_2017}.
Thus, the particle-scale mechanism of shear thickening 
is, to a great extent, understood.


However, particle-scale simulations are not capable of reproducing engineering-scale 
flows of dense suspensions due to the practical limits imposed on the system size by computational tractability.
For this reason, it is important to develop effective continuum models through the design of suitable non-Newtonian constitutive relations.
Besides laboratory experiments, particle-scale simulations are an important source of indications for the development of such models.
A complete model should describe the fluid response under any flow condition~\citep{Miller_2009}, 
not only in the simple shear flows in which most experimental and computational data are retrieved.
Indeed, the response of non-Newtonian fluids can depend on the type of flow, as exemplified by the observations 
of shear thinning and extensional thickening in some viscoelastic fluid.
Particularly important is the class of extensional flows of suspensions, 
for which few rheological characterisations are available~\citep{Dai_2017} 
and the sole computational investigation of which the authors are aware 
was performed by \citet{Sami_1996},
who studied semidilute Brownian suspensions.
(We note that in his analysis flow-type dependence was not evidenced.)
A related computational method to treat hydrodynamic interactions
in diluted suspensions was introduced by \citet{Ahamadi_2008}.
For important developments regarding emulsions of deformable droplets,
we refer the reader to the work of \citet{Zinchenko_2015}.

To study the material response,
we simulate motions of particles in the bulk region under prescribed flow conditions.
As usual, periodic boundary conditions are employed to minimise finite-size effects.
The Lees--Edwards boundary conditions~\citep{Lees_1972}
are commonly used to impose simple shear flows in many contexts, including suspension rheology~\citep{Bossis_1984,Mari_2014}.
In this work, we also apply the Kraynik--Reinelt 
boundary conditions~\citep{Kraynik_1992,Todd_1998}, 
originally devised 
to impose planar extensional flows in nonequilibrium molecular dynamics simulations.
With these we can provide a first assessment of the flow-type dependence 
of the response in dense suspensions.


In \secref{sec_pbc} and \secref{sec_particle_dynamics} we describe our simulation technique which operates in the inertialess approximation.
To compare consistently the results under different flow conditions,
we employ the rheometric framework introduced by \citet{Giusteri_2017} 
(summarised in \secref{sec_general_response_functions}) which defines, 
for the case of planar flows, a dissipative response function, $\kappa$, 
and two non-dissipative response functions, $\lambda_0$ and $\lambda_3$.
Those are defined for any flow type (simple shear, extensional, and mixed flows) 
and offer a unified description of the material response.
The results of our analysis, discussed in \secref{sec_results}, 
highlight the presence of flow-type dependence in the microstructure 
and in the non-Newtonian effects observed for dense suspensions.

\section{Methods}

\subsection{Bulk rheology with periodic boundary conditions}
\label{sec_pbc}

Non-Newtonian incompressible fluids obey the differential equations
\begin{equation}
  \rho 
\left[
    \frac{\partial \bm{u}}{\partial t}
    + (\bm{u} \cdot \nabla)\bm{u}
\right]
  = \nabla \cdot \CauchyStress
  \qquad\text{ with }\qquad
  \nabla \cdot \bm{u}=0,
  \label{eq_Cauchy_momentum_eq}
\end{equation}
where $\bm{u}$ is the velocity field 
and $\rho$ the density.
To close the system of equations, the stress tensor $\CauchyStress$
must be given in terms of the velocity gradient through a constitutive prescription.
The local value of the stress tensor describes the material response and 
is determined by the local history of deformation.
To investigate such response,
we consider small volume elements
in which the velocity gradient $\nabla \bm{u}$ is approximately uniform.
By simulating motions of particles in the volume element 
with fixed $\nabla \bm{u}$,
we can find the typical stress for a certain deformation history.

\begin{figure}
  \centerline{\includegraphics[width=0.98\columnwidth]{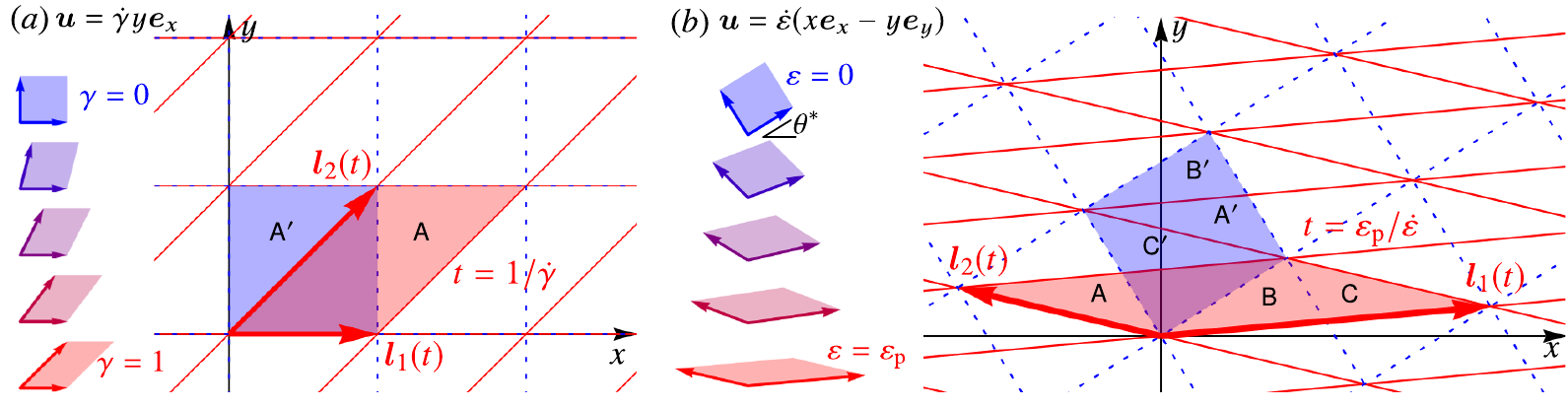}}
\caption{
(\textit{a})~The deforming periodic cells for simple shear flows.
The initial rectangular shape can be recovered when shear strain $\gamma$ equals $1$
by removing a part $\mathsf{A}$ of the master cell
and including the corresponding part $\mathsf{A}'$
of a periodic image in the new master cell.
Note that the recovery can be performed at 
any value of the shear strain $\gamma$ if 
the periodic displacement of the rows is taken into account.
(\textit{b})~The deforming periodic cells for extensional flow
(Kraynik--Reinelt boundary conditions).
The initial rectangular cell,
which is oriented at a certain angle $\theta^{\ast} \approx \SI{31.7}{\degree}$,
can be recovered when the strain $ \varepsilon$ 
equals $\varepsilon_{\mathrm{p}} \approx 0.962 $
by removing parts $\mathsf{A}$--$\mathsf{C}$ of the master cell
and including the corresponding parts 
$\mathsf{A}'$--$\mathsf{C}'$
of periodic images in the new master cell.
}
\label{fig_pbc}
\end{figure}

Time-dependent periodic boundary conditions allow to impose $\nabla \bm{u}$ 
and effectively simulate the bulk behaviour.
Since we will consider planar flows in a 3D geometry, we can describe our methods
considering the 2D projections of the computational cells.
The cell frame vectors $\bm{l}_1(t)$ and $\bm{l}_2(t)$ (see \figref{fig_pbc})
are prescribed to follow the velocity field $\bm{u}=\nabla \bm{u} \cdot \bm{r}$
and periodic images of a particle at $\bm{r}$
are given by $\bm{r}' = \bm{r} + i \bm{l}_1(t)+ j \bm{l}_2(t)$ with ($i,j = \pm 1, \pm 2, \dotsc$).
For simple shear flows ($\nabla \bm{u}=\dot{\gamma} \bm{e}_y \bm{e}_x$),
this is equivalent to the Lees--Edwards boundary conditions.
The initial periodic cells are rectangles in the flow plane
(blue in \figref{fig_pbc}\,(\textit{a})).
A simple shear flow
deforms the cells to parallelogram shapes
(red in \figref{fig_pbc}\,(\textit{a})).
To avoid significantly deformed periodic cells,
the initial rectangular cells can be recovered
as shown in \figref{fig_pbc}\,(\textit{a}).
To impose planar extensional flows 
($\nabla\bm{u} =\dot{\varepsilon} \bm{e}_x \bm{e}_x - \dot{\varepsilon}\bm{e}_y \bm{e}_y$)
for long times, 
we employ the Kraynik--Reinelt periodic boundary conditions~\citep{Kraynik_1992,Todd_1998}.
If the initial master cell is a regular square 
oriented at a certain angle $\theta^{\ast}$ from 
the extension axis ($x$-axis),
the deformed parallelogram cell 
after a certain strain $\varepsilon_{\mathrm{p}}$ 
can be remapped to the initial regular shape 
as shown in \figref{fig_pbc}~(\textit{b}).

\subsection{Inertialess particle dynamics for suspensions}
\label{sec_particle_dynamics}

We numerically evaluate the stress tensor $\CauchyStress$ 
by using particle simulations with deforming periodic cells.
Our simulation is analogous to rate-controlled rheological measurements 
in the sense that 
time-averaged stress responses $\langle \CauchyStress\rangle$ 
are evaluated for imposed velocity gradients $\nabla \bm{u}$.
%

We consider non-Brownian, density matched, and dense suspensions.
Suspended particles interact with each other in several ways.
As discussed in \citet{Mari_2014},
we take into account contact forces $\bm{F}_{\mathrm{C}}$ (and torques $\bm{T}_{\mathrm{C}}$)
and stabilising repulsive forces $\bm{F}_{\mathrm{R}}$,
besides hydrodynamic interactions $\bm{F}_{\mathrm{H}}$ and $\bm{T}_{\mathrm{H}}$.
Since the inertia of sufficiently small particles
is negligible in comparison to the hydrodynamic drag forces,
the particles obey the quasi-static equations of motion
\begin{equation}
  \bm{F}_{\mathrm{H}} + \bm{F}_{\mathrm{C}} + \bm{F}_{\mathrm{R}}  
  =  \bm{0}
\quad\text{and}\quad
  \bm{T}_{\mathrm{H}} + \bm{T}_{\mathrm{C}} = \bm{0}.
  \label{eq_force_balance_h_c_r}
\end{equation}
Here, forces $\bm{F}$ and torques $\bm{T}$ 
represent the set of forces and torques for $N$ particles. 
Flows around microscale particles
are dominated by viscous dissipation and the inertia of the fluid is negligible,
so they are described by the Stokes equations.
The imposed velocity gradient $\nabla \bm{u}$
gives the background flow
via
the velocity $\bm{u}(\bm{r})$,
vorticity $\bm{\omega} \equiv \nabla \times \bm{u}$,
and rate of deformation tensor $\Tensor{D}$ such that
$\nabla \bm{u} \cdot \bm{r} 
=
\bm{u}(\bm{r})  
= \Tensor{D} \cdot \bm{r}
+ (\bm{\omega}/2) \times \bm{r}.$
In this case, the hydrodynamic interactions can be expressed as the sum 
of linear resistances to the relative velocities 
$\varDelta \bm{U}^{(i)} \equiv \bm{U}^{(i)}-\bm{u}(\bm{r}^{(i)})$, 
angular velocities 
$\varDelta \bm{\varOmega}^{(i)} \equiv\bm{\varOmega}^{(i)}-\bm{\omega}/2$, for $i = 1, \dotsc, N$, 
and imposed deformation $\Tensor{D}$ via
\begin{equation}
  \begin{pmatrix}
    \bm{F}_{\mathrm{H}} \\ 
    \bm{T}_{\mathrm{H}} 
  \end{pmatrix}
  = 
  -\Tensor{R} \cdot 
  \begin{pmatrix}
    \varDelta \bm{U} \\
    \varDelta \bm{\varOmega}\\
  \end{pmatrix}
  +\Tensor{R}'
  \cddot \Tensor{D}_{N}
,\label{eq_hydro_force}
\end{equation}
where $\varDelta\bm{U}$ and $\varDelta \bm{\varOmega}$
represent the set of relative velocities for $N$ particles,
$\Tensor{D}_{N}$ is block-diagonal with $N$ copies of $\Tensor{D}$,
and $\Tensor{R}$ and $\Tensor{R}'$ are resistance matrices
which can, in principle, be derived from the Stokes equations 
once the particle configurations are given.
In dense suspensions,
the long-range hydrodynamic interactions are screened by crowds of particles.
Therefore, we may approximately construct the resistance matrices
by including only the contributions of Stokes drag and lubrication forces.
%


In real suspensions, the lubrication singularity in $\bm{F}_{\mathrm{H}}$
is absent due to factors such as the surface roughness of particles---direct contacts are not forbidden.
Hence, we include the contact interactions $\bm{F}_{\mathrm{C}}$ and $\bm{T}_{\mathrm{C}}$
in \eqref{eq_force_balance_h_c_r}.
The contact forces between solid particles 
depend on the nature of the particle surfaces.
This is effectively encoded in the friction coefficient $\mu$
that enters a simple friction model.
By denoting with
$F_{\mathrm{C}}^{n}$ and $F_{\mathrm{C}}^{t}$
normal and tangential forces, respectively,
we prevent sliding if 
$F_{\mathrm{C}}^{t} \leq \mu F_{\mathrm{C}}^{n}$.
The normal force depends on the overlap between particles
through an effective elastic constant 
and the tangential force depends on the sliding displacement in a similar way.
The details of the employed model are given in \citet{Mari_2014}.
%

%

The presence of the stabilising repulsive force $\bm{F}_{\mathrm{R}}$ in \eqref{eq_force_balance_h_c_r}
generates the rate dependence of rheological properties in such suspensions.
Indeed, while reaching the same strain, $\bm{F}_{\mathrm{R}}$ can work more 
to prevent particle contacts 
under lower deformation rates, 
but less under higher rates. 
As a result, the number of contacts depends on the rate of the imposed flow.
In colloidal suspensions,
Brownian forces may play a similar role, as discussed by \citet{Mari_2015a}.
%


The bulk stress tensor is obtained as
\begin{equation}
  \CauchyStress 
  = 
  -p_0 \Tensor{I}
  +
  2 \eta_0 \Tensor{D}
  + 
  V^{-1}
\biggl(
\sum_i \Tensor{S}_{\mathrm{D}}^{(i)}
+ 
\sum_{i > j} 
\Tensor{S}_{\mathrm{P}}^{(i,j)}
\biggr),
  \label{eq_part-mot}
\end{equation}
where 
$\eta_0$ is the viscosity of the solvent,
$\Tensor{S}_{\mathrm{D}}^{(i)}$
is the stresslet on particle $i$ due to $\Tensor{D}$,
and $\Tensor{S}_{\mathrm{P}}^{(i,j)}$
is the stresslet due to non-hydrodynamic interparticle forces 
between particle $i$ and $j$~\citep{Mari_2015}.
Note that, since the hydrostatic pressure $p_0$ is arbitrary, we set $p_0=0$.
However, the last term in \eqref{eq_part-mot} is not traceless and 
thus contributes to the total pressure $p \equiv -(1/3)\tr{\CauchyStress}$.

\subsection{General response functions for steady flows of non-Newtonian fluids}
\label{sec_general_response_functions}

The stress $\CauchyStress$ is a tensorial quantity 
and we need a procedure to extract from 
it the relevant information in terms of scalar quantities.
We are interested in comparing the material response 
under different types of imposed flow conditions.
For this reason, we need a framework in which
it is possible to identify the dependence on the flow type of
each independent non-Newtonian effect.
To this end, we use the framework introduced by \citet{Giusteri_2017},
in which the characteristic rate of the imposed flow
is defined independently of the flow type
and a complete set of response functions is given.
These functions generalise to any flow type standard quantities
such as viscosity and normal stress differences.


%
The velocity gradient $\nabla \bm{u}$ is decomposed 
into symmetric and antisymmetric parts as
$\nabla \bm{u} = \Tensor{D} + \Tensor{W}$.
In the planar case, with $\Tensor{D} \neq \Tensor{0}$, 
we denote by $\dot{\varepsilon}$ the largest eigenvalue of $\Tensor{D}$ 
and express $\Tensor{\hat{D}}  \equiv \Tensor{D}/ \dot{\varepsilon} $ 
and $\Tensor{\hat{W}} \equiv \Tensor{W} /\dot{\varepsilon} $ 
on the basis of the eigenvectors 
$\hat{\bm{d}}_1$ and $\hat{\bm{d}}_2$ of $\Tensor{D}$ 
(corresponding to the eigenvalues $\dot{\varepsilon}$ and $-\dot{\varepsilon}$) as follows:
\begin{equation}
  \Tensor{\hat{D}}  = 
  \hat{\bm{d}}_1\hat{\bm{d}}_1 - \hat{\bm{d}}_2\hat{\bm{d}}_2,
  \quad
  \Tensor{\hat{W}}
  = 
  \beta_3
  (\hat{\bm{d}}_2\hat{\bm{d}}_1 - \hat{\bm{d}}_1\hat{\bm{d}}_2).
  \label{eq_D_and_W}
\end{equation}
The non-vanishing and positive rate $\dot{\varepsilon} > 0$
is used to set the time scale of deformation in any flow type.
With this definition,
the standard rate $\dot{\gamma} $ for simple shear
corresponds to the value $2\dot{\varepsilon}$.
The vorticity $\omega_z$ 
is represented by the dimensionless parameter $\beta_3$ 
through $\omega_z = 2 \dot{\varepsilon} \beta_3$. 
Note that planar extensional flows are characterised by $\beta_3 = 0$,
and simple shear flows by $\beta_3 = 1$.
%


A general representation of the stress tensor in planar flows is then given by 
\begin{equation}
  \CauchyStress (\dot{\varepsilon}, \beta_3)
= - p(\dot{\varepsilon}, \beta_3)
\Tensor{I}  +  
\dot{\varepsilon}
\bigl[
\kappa (\dot{\varepsilon}, \beta_3) \Tensor{\hat D} 
+
\lambda_0 (\dot{\varepsilon}, \beta_3) \Tensor{\hat E}
+
\lambda_3(\dot{\varepsilon}, \beta_3) \Tensor{\hat G}_3
\bigr],\label{133900_27Jun17}
\end{equation}
where 
$\Tensor{\hat E} \equiv
-(1/2)(\hat{\bm{d}}_1\hat{\bm{d}}_1+\hat{\bm{d}}_2\hat{\bm{d}}_2)
 +\hat{\bm{d}}_3\hat{\bm{d}}_3 $, $\hat{\bm{d}}_3$ is 
the eigenvector of $\Tensor{D}$ orthogonal to the flow plane,
and 
$ \Tensor{\hat G}_3 \equiv 
\hat{\bm{d}}_1\hat{\bm{d}}_2 + \hat{\bm{d}}_2\hat{\bm{d}}_1 $
is introduced to complete 
an orthogonal basis for the space of symmetric tensors for planar flows.
The functional dependence of
$\kappa$, $\lambda_0$, and $\lambda_3$ 
on the two kinematical parameters $\dot{\varepsilon}$ and  $\beta_3$
needs to be determined to characterise the response in generic flows.
We remark that the response functions $\kappa$, $\lambda_0$, and $\lambda_3$ 
can depend on any other quantity that characterise the system.
For instance, in \secref{sec_results} 
we will also show their dependence on the volume fraction $\phi$.
The function $\kappa$ is the only one to carry information about dissipation.
We therefore refer to it as the \emph{dissipative response function},
a generalised viscosity.
The functions $\lambda_0$ and $\lambda_3$ 
carry information about non-dissipative responses 
and we call them \emph{non-dissipative response functions}.
The presence of a nonvanishing $\lambda_0$ leaves the eigenvectors 
of the stress $\CauchyStress$ aligned with those of $\Tensor{D}$,
as happens in Newtonian fluids, 
but gives a contribution to the stress in the form of a modified pressure 
which is isotropic in the flow plane but different in the direction normal to the flow plane.
On the other hand, a nonvanishing $\lambda_3$ 
corresponds to a rotation of the eigenvectors of $\CauchyStress$ 
in the flow plane with respect to those of $\Tensor{D}$, 
determining the reorientation angle
\begin{equation}
 \varphi \equiv \arctan\Bigl[
  \lambda_3 \Big/ \Bigl(\kappa + \sqrt{ \kappa^2 +\lambda_3^2 }\Bigr)
  \Bigr].
\label{eq:reorientation_angle}
\end{equation}

For the sake of comparison,
the shear viscosity $\eta$ 
and normal stress differences $N_1$ and $N_2$,
defined for simple shear flows with $\beta_3 = 1$
as functions of $\dot{\gamma} = 2 \dot{\varepsilon}$,
are given  by 
\begin{equation}
\eta(2\dot{\varepsilon})=\kappa(\dot{\varepsilon},1)/2, \quad
N_1(2\dot{\varepsilon})
=-2\dot{\varepsilon}\lambda_3(\dot{\varepsilon},1),
\quad\text{and}\quad
N_2(2\dot{\varepsilon})=\dot{\varepsilon}
[\lambda_3(\dot{\varepsilon},1)
-(3/2)\lambda_0(\dot{\varepsilon},1)].
\end{equation}
Moreover, the extensional viscosity,
defined for extensional flows with $\beta_3 = 0$,
is given by 
$\eta_{\mathrm{E}}(\dot{\varepsilon}) =  2 \kappa(\dot{\varepsilon}, 0)$.
So that the Trouton ratio 
$\eta_{\mathrm{E}}(\dot{\varepsilon})/
\eta(2\dot{\varepsilon}) $ equals $4$
only if $\kappa(\dot{\varepsilon},0) =\kappa(\dot{\varepsilon},1)$.


%
We want to stress that,
to arrive at \eqref{133900_27Jun17},
no a priori assumption 
is made on the list of quantities
on which the material functions can depend.
Hence, the stress tensor for any non-Newtonian fluid model
under steady flow conditions can be expressed
in the form \eqref{133900_27Jun17}.
For example, 
since $\Tensor{D}^2$ in planar flows 
is a linear combination of $\Tensor{I}$ and $\Tensor{\hat{E}}$,
the class of Reiner--Rivlin fluids
corresponds to choosing $\lambda_3 = 0$,
and assuming $\kappa$ and $\lambda_0$ 
independent of $\beta_3$.
Similarly, second-order fluids under steady shear flows
would produce constant values of $\kappa$ and $\lambda_0$, 
and entail $\lambda_3 \propto \beta_3 \dot{\varepsilon}$,
since, under such flows, 
$\Tensor{W}\cdot\Tensor{D}-\Tensor{D}\cdot\Tensor{W}
= \beta_3 \dot{\varepsilon}^2 \Tensor{\hat{G}}_3$.
A detailed discussion of the representation \eqref{133900_27Jun17}
and its relation to fluid models are given in \citet{Giusteri_2017}.

\section{Results}
\label{sec_results}

To obtain the numerical results, we mainly performed 50 independent 3D simulations with $2000$ particles.
The periodic cells are initially cuboids with ratio $5:5:1$.
We also performed some simulations with 4000 particles using double-sized cells ($5:5:2$) 
to confirm the absence of significant finite-size effects (data are not shown).
Regarding the friction coefficient, we set $\mu = 1$ since it is the value that,
in a previous paper \citep{Mari_2015a}, was found to give 
good agreement with the experimental data by \citet{Cwalina_2014}.
It is worth mentioning that \citet{Tanner_2016}
showed that $\mu = 0.5$ gives a better quantitative agreement 
with different experimental data.
Nevertheless, such a fine tuning of $\mu$ is not necessary
for our qualitative analysis.


%
The short-range repulsive force
is given by $|\bm{F}_{\mathrm{R}}| = F_{\mathrm{R}}(0) \exp[- (r-2a)/(0.02a)]$,
with $a$ the particle radius.
A reference rate is set as $\dot{\varepsilon}_0 \equiv F_{\mathrm{R}}(0) / (12 \pi \eta_0 a^2)$.
To estimate the importance of the inertial effects, 
we can use the Stokes number given by
\begin{equation}
\StNum
\equiv \frac{2 \rho_{\mathrm{p}} a^2 \dot{\varepsilon}}{\eta_0}
= \frac{\rho_{\mathrm{p}} F_{\mathrm{R}}(0)}{6\pi \eta_0^2}
\frac{\dot{\varepsilon}}{\dot{\varepsilon}_0},
\end{equation}
with $\rho_{\mathrm{p}}$ the particle density.
Hence, inertial effects can be neglected if $\mathrm{St}\ll 1$, that is
when the ratio $\dot{\varepsilon}/\dot{\varepsilon}_0$
is much smaller than $6\pi \eta_0^2 / \rho_{\mathrm{p}} F_{\mathrm{R}}(0)$.
The preceding threshold determines, for each specific system, 
the region in which the rheology curves obtained with our simulations 
can be expected to be in agreement with real data.

\subsection{Dissipative response function $\kappa$}
\label{sec_kappa}

\begin{figure}
\centerline{\includegraphics[width=0.45\textwidth]{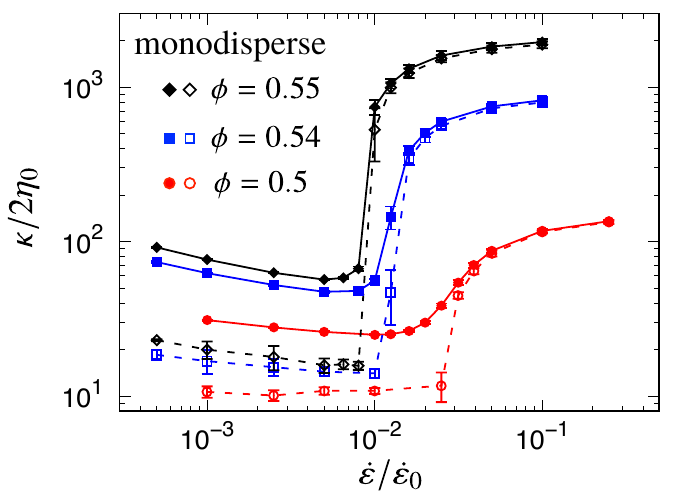}}
\caption{
Both shear thickening and extensional thickening are observed
in the rate dependence of the dissipative material function $\kappa$.
Data are for monodisperse suspensions
in extensional flows (filled symbols with solid lines)
and
in simple shear flows (open symbols with dashed lines).
Below thickening, $\kappa$ in simple shear flows 
is much lower than $\kappa$ in extensional flows.
In each simulation, the time average is taken over 5 strains after reaching the steady state.
The error bars show standard deviation for 50 independent simulations.
}
\label{fig_kappa_mono}
\end{figure}

For the case of monodisperse suspensions,
the dissipative response function $\kappa$
significantly increases with the rate $\dot{\varepsilon}$ 
both in simple shear and extensional flows (\figref{fig_kappa_mono}).
Not only shear thickening but also extensional thickening occurs.
However, below thickening,
there is a clear flow-type dependence.
The value of $\kappa$ in extensional flow is 
much higher than the one obtained in simple shear flow 
(the Trouton ratio is much larger than $4$).
On the other hand,
above thickening,
the values of $\kappa$ in extensional and shear flows
are almost indistinguishable (the Trouton ratio is very close to $4$)
and the flow-type dependence is hindered.


The significant discrepancy observed below thickening is due to shear-induced ordering,
which can occur only in simple shear flows, 
as we confirm by analysing the pair distribution functions in \secref{sec_microstructure}.
Since streamlines of a simple shear flow 
are straight and parallel to each other,
particles tend to be arranged in chain-like structures along the flow direction.
We observe a gradual decrease of $\kappa$ over time (strain thinning) 
in simple shear flows,
which indicates the growth of the ordered structure. 
It should be noted that
the shear-induced ordering is enhanced by the periodic boundary conditions,
since linear chains may connect with their own periodic images.
By contrast, the streamlines of extensional flows
are never parallel to each other.
Therefore, there is no obvious ordered structure compatible with extensional flows.
Indeed, we neither observe strain thinning 
nor any ordered microstructure in the extensional flow simulation.

\begin{figure}
\centerline{\includegraphics[width=0.9\textwidth]{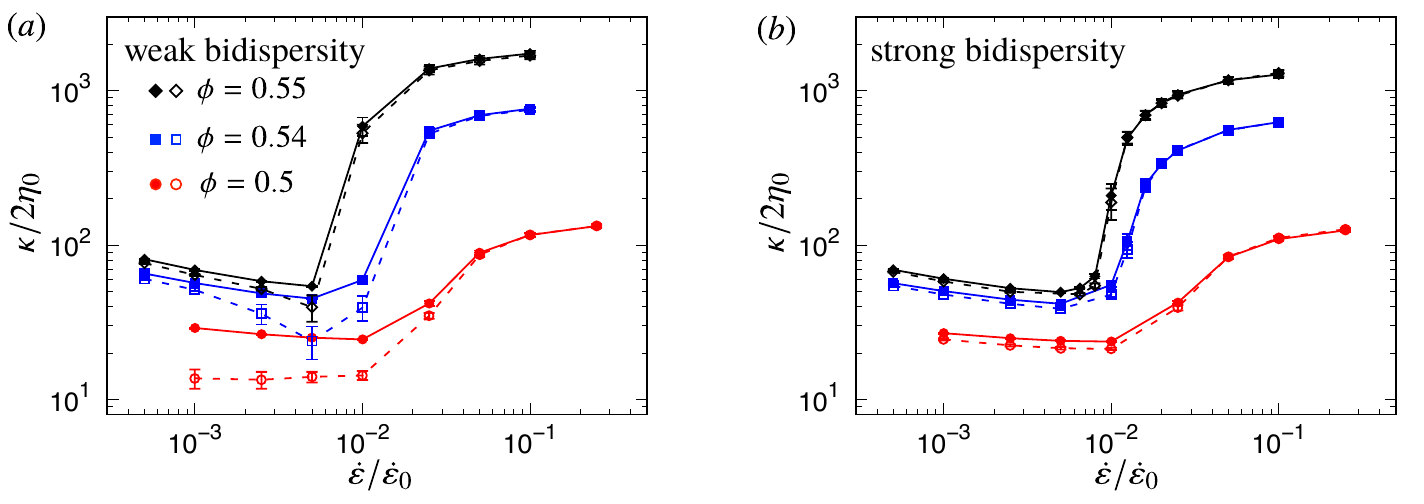}}
\caption{
Mixing particles of different sizes
hinders the shear-induced ordering.
Differences in $\kappa$ between extensional flow (filled symbols with solid lines) 
and simple shear (open symbols with dashed lines)
are still present in the weakly bidisperse suspensions, $a_2/a_1 = 1.2$ (\textit{a}),
but no longer significant in the strongly bidisperse suspensions, $a_2/a_1 = 1.4$ (\textit{b}).
}
\label{fig_kappa_bi}
\end{figure}

In the thickened regime,
frictional contact forces are constantly activated.
Such contact forces are so strong that 
particles are easily prevented from following the background flow,
thus ordered structures cannot be developed.
As long as the disordered structure is maintained under simple shear flows,
the value of $\kappa$ remains very close to that observed in extensional flows.


The shear-induced ordering can be hindered by mixing particles with different sizes.
To see this effect, we consider two types of bidisperse suspensions 
with different size ratios: $a_2/a_1 = 1.2$ and $a_2/a_1 = 1.4$
(named ``weak'' and ``strong'', respectively).
Two populations occupy the same volume fractions,
i.e., $\phi_1 = \phi_2 = \phi /2$. 
In the weakly bidisperse suspensions (\figref{fig_kappa_bi}\,(\textit{a})),
although the differences clearly become smaller,
some flow-type dependence can still be seen, especially for $\phi = 0.5$.
In the strongly bidisperse suspensions (\figref{fig_kappa_bi}\,(\textit{b})),
we no longer see a noticeable flow-type dependence---Trouton ratios are always close to $4$.


\subsection{Pressure and anisotropic response}
\label{sec_lambda0}

\begin{figure}
\centerline{\includegraphics[width=0.9\columnwidth]{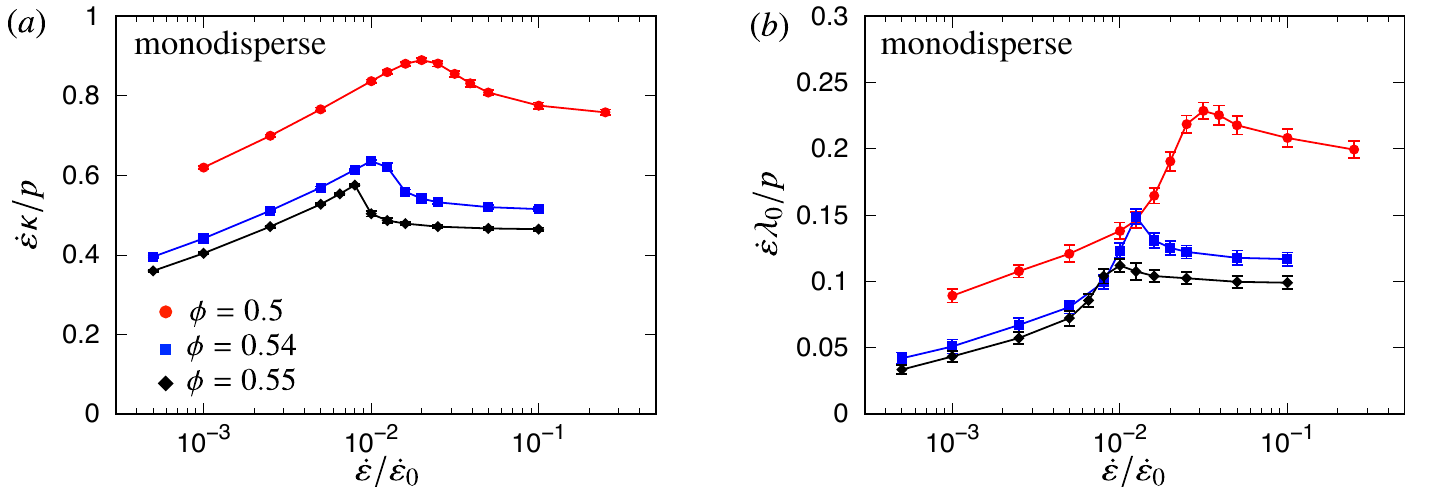}}
\caption{
(\textit{a})
The ratio between $\dot{\varepsilon}\kappa$ and $p$ 
for monodisperse suspensions in extensional flows
remains of the order of unity even when the viscosity increases significantly.
(\textit{b})
The positive values of the ratio $\dot{\varepsilon} \lambda_0/p$
for monodisperse suspensions in extensional flows
indicate some anisotropy in the pressure response,
namely the in-plane pressure is higher than the out-of-plane pressure.
}
\label{fig_pressure_and_lambda0}
\end{figure}

The total stress tensor $\CauchyStress$ is usually split into two parts: 
isotropic pressure term and traceless extra-stress term.
Though only the extra-stress term determines the flows of incompressible fluids,
the pressure $p$ is also a part of the material response.
As seen in \figref{fig_pressure_and_lambda0}\,(\textit{a})
for monodisperse suspensions in extensional flows,
the pressure term $p$ varies in a similar way as $\dot{\varepsilon} \kappa$;
the ratio $ \dot{\varepsilon} \kappa / p $ 
remains of the order of unity even when $\kappa$ significantly increases by thickening.
In our simulation, the volume of the periodic cells is fixed,
therefore the system can never dilate.
However, such increase of $p$ with $\dot{\varepsilon} \kappa$
suggests that extensional thickening (and shear thickening) of suspensions 
is a phenomenon related to that of \emph{dilatancy} in granular materials.


The pressure term $p$ contributes isotropically to $\CauchyStress$ by definition.
However, 
there is another contribution to the stress $\CauchyStress$
sharing the same origin.
The non-dissipative response associated with dilatancy
can be anisotropic and activate the response function $\lambda_0$.
The dimensionless ratio $\dot{\varepsilon} \lambda_0/p$
represents such anisotropy.
Its positive values reported in \figref{fig_pressure_and_lambda0}\,(\textit{b})
indicate that the in-plane pressure 
is higher than the out-of-plane pressure.
However, this anisotropy is not very strong, 
as it would be if the pressure dilatancy were only present in the flow plane.

\subsection{Reorientation angle of stress eigenvectors}
\label{sec_lambda3}

\begin{figure}
\centerline{\includegraphics[width=\textwidth]{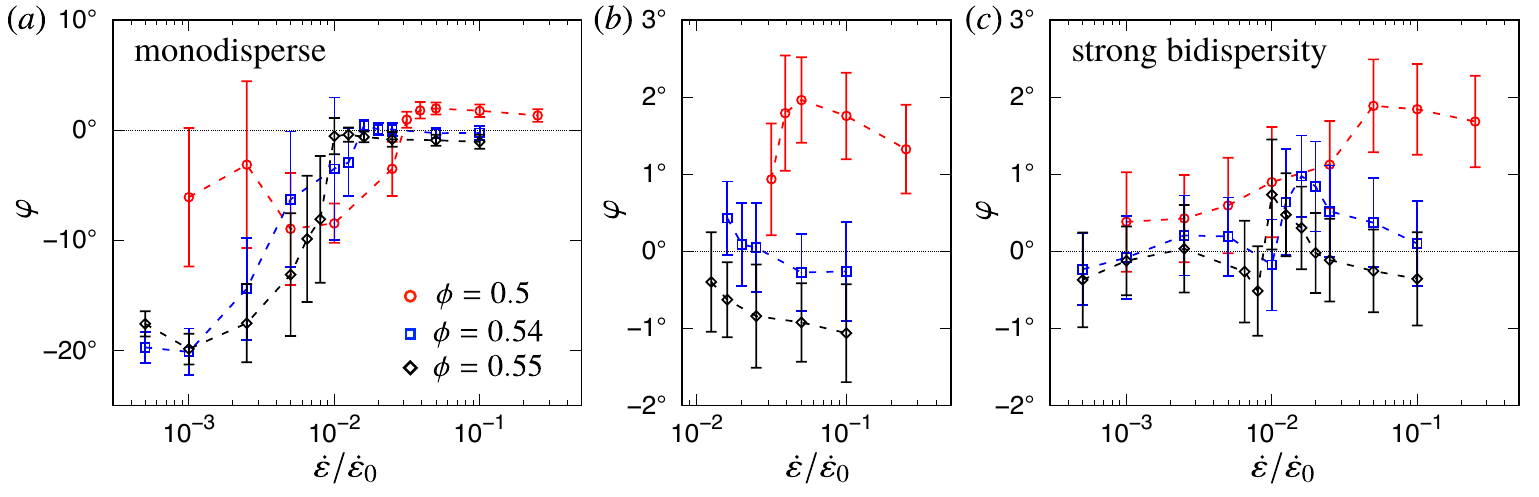}}
\caption{
The reorientation angle $\varphi$
is non-zero in simple shear flows
for both monodisperse suspensions (\textit{a})--(\textit{b}) 
and strongly bidisperse suspensions (\textit{c}),
while it fluctuates around zero in extensional flows (not shown).
Such angle is associated with the first normal stress difference $N_1$
and we have 
$\varphi \approx - N_1 / (4\dot{\varepsilon} \kappa) $ when $\lambda_3 \ll \kappa$.
The large standard deviations 
present in the monodisperse case (\textit{a}) below thickening
are due to the existence of several types of stable ordered structures
displaying rather different values of $\varphi$.
This is likely to be an effect originated by the finite-size of the simulation cell.
When the microstructure is disordered (\textit{b})--(\textit{c}),
the standard deviations are smaller and comparable.
}
\label{fig_lambda3}
\end{figure}

Besides the ordering in simple shear flow,
we can see some flow-type dependence 
in the reorientation angle~$\varphi$,
defined in \eqref{eq:reorientation_angle}.
In extensional flows, 
the principal axes of the stress tensor $\CauchyStress$
must be parallel to
the eigenvectors of $\Tensor{D}$ due to symmetry considerations.
Indeed, the reorientation angle $\varphi$ fluctuates around zero in those simulations.
In simple shear flows,
the shear-induced ordering 
is accompanied by large negative values of $\varphi$~(\figref{fig_lambda3}~(a)).
On the other hand, in the disordered states 
above thickening~(\figref{fig_lambda3}~(b))
and with strong bidispersity~(\figref{fig_lambda3}~(c)),
$\varphi$ is always rather small but non-zero.
In our inertialess simulation,
this finite flow-type dependence indicates
some characteristic microstructure 
(see \secref{sec_microstructure})
due to the presence of vorticity in simple shear,
which is absent in extensional flows.
%


It is worth commenting on the dependence of the angle $\varphi$ 
on the volume fraction $\phi$ (\figref{fig_lambda3}).
In the thickened regime,
corresponding to higher values of $\dot{\varepsilon}$,
the angle $\varphi$ is always positive for $\phi = 0.5$.
The values of $\varphi$ become smaller and can take slightly negative values
as $\phi$ increases.
This behaviour is consistent with some experimental measurements of $N_1$, 
which is proportional to $-\lambda_3$.
When the volume fraction is not very high,
negative values have 
been observed for $N_1$~\citep{Lee_2006a,Cwalina_2014}, corresponding to positive $\varphi$,
while the sign of $N_1$ turns positive (negative $\varphi$)
at higher volume fractions~\citep{Lootens_2005,Dbouk_2013}.

\subsection{Microstructure}
\label{sec_microstructure}

\begin{figure}
\centerline{\includegraphics[width=1\textwidth]{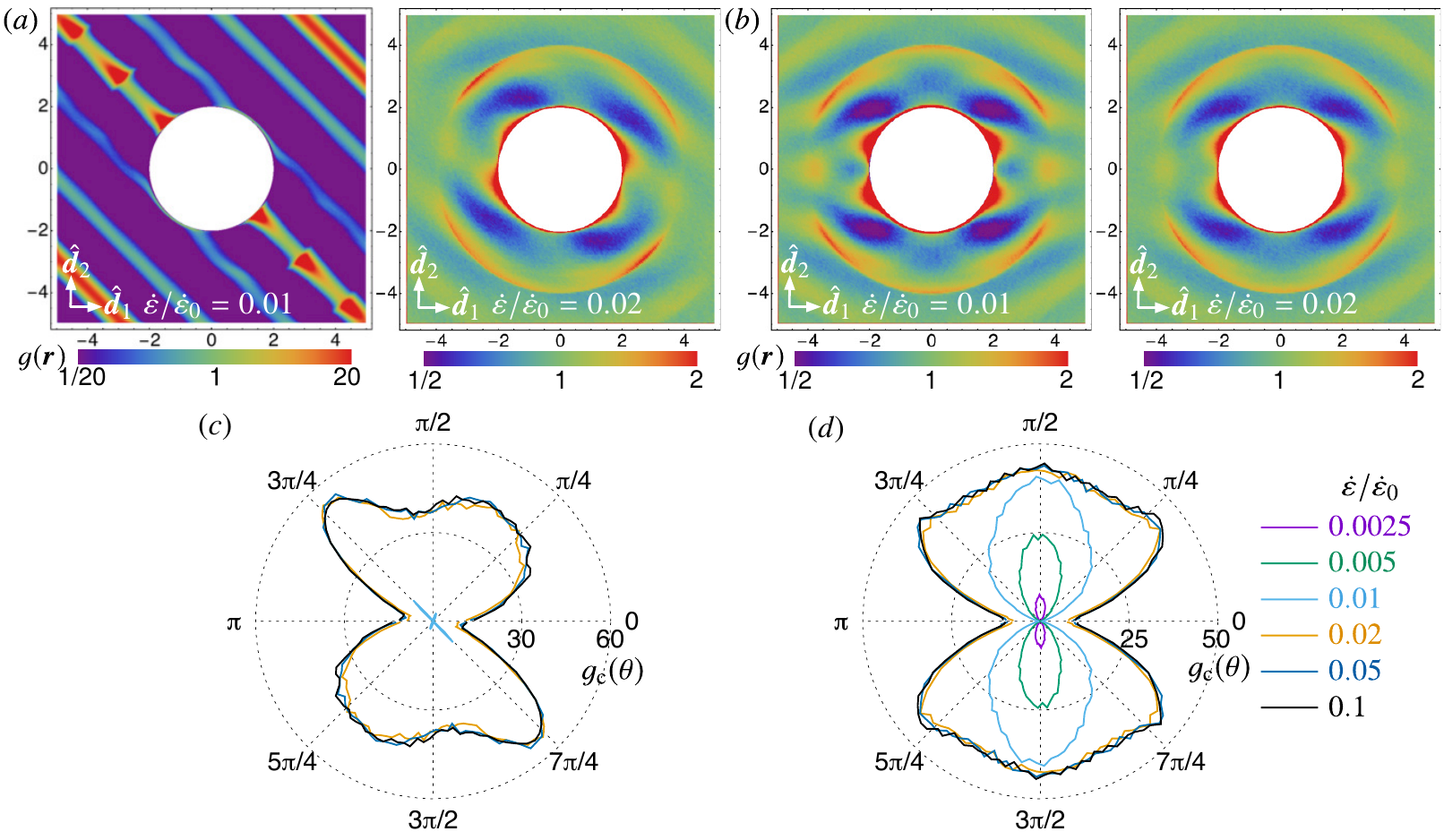}}
\caption{
(\textit{a})
The pair distribution function $g(\bm{r})$ 
highlights the presence of strong ordering
below thickening ($\dot{\varepsilon}/\dot{\varepsilon}_0 = 0.01$)
for monodisperse suspensions ($\phi = 0.54$)
under simple shear, and its absence above thickening~($\dot{\varepsilon}/\dot{\varepsilon}_0 = 0.02$).
(\textit{b})
No obvious ordered structure
can be associated to $g(\bm{r})$ 
for the case of extensional flow
both below and above thickening.
(\textit{c})--(\textit{d})
The polar plot of the angular distribution $g_{\mathrm{c}}(\theta)$ 
of contacting (and nearly contacting) particles such that $|\bm{r}| < 2.02 a$
shows that the strong enhancement
of contact interactions is the main responsible of thickening.
Note that, 
since practically no particles are in contact 
for $\dot{\varepsilon}/\dot{\varepsilon}_0 \le 0.005$
in simple shear, 
the corresponding data in (\textit{c}) are negligible.
The difference between simple shear flow (\textit{c}) and extensional flow (\textit{d})
is discussed in the main text.
}
\label{fig_microstructure}
\end{figure}

As discussed in the modelling section \secref{sec_particle_dynamics},
it is reasonable to neglect particle and fluid inertia
in the particle-scale dynamics.
The response of such inertialess material elements to an imposed flow 
essentially depends on the microstructure 
built by the particles during the flow~\citep{Morris_2009}.
To measure the correlation of particle positions,
we evaluate the pair distribution function $g(\bm{r}) \equiv P_{1|1}(\bm{r}|\bm{0})/n$,
where $n$ is the average number density of particles
and $P_{1|1}(\bm{r}|\bm{0})$
is the conditional probability of finding a particle at $\bm{r}$
with the condition that another particle is at the origin $\bm{0}$.
\figref{fig_microstructure}\,(a) and (b) show 
$g(\bm{r})$ in the flow-plane slice $|z| < 0.1 a$.
We also consider angular distributions $g_{\mathrm{c}}(\theta)$ 
for contacting (and nearly contacting) particles such that $|\bm{r}| < 2.02 a$.
The angle $\theta$ is measured from the $\hat{\bm{d}}_1$ axis.


As seen in \figref{fig_microstructure}\,(\textit{a}),
a stripe-patterned correlation $g(\bm{r})$ appears 
for the monodisperse suspensions in simple shear flow
below thickening ($\dot{\varepsilon}/\dot{\varepsilon}_0 = 0.01$).
The periodic peaks and striped correlation indicate 
the formation of chain-like structures by the particles.
Once such chain-like structure is formed, 
particle interactions are rather weak,
which leads to significantly low values of $\kappa$ 
as seen in \figref{fig_kappa_mono}.
The microstructure is totally different 
above thickening ($\dot{\varepsilon}/\dot{\varepsilon}_0 = 0.02$).
The long-range correlation is no longer seen.
The correlation pattern indicates some disordered anisotropic microstructure.
In \figref{fig_microstructure}\,(\textit{c}), 
the angular contact distribution $g_{\mathrm{c}}(\theta)$ 
clearly shows that the number of contacting particles remarkably increases
for $\dot{\varepsilon}/\dot{\varepsilon}_0 \geq 0.02$.
This observation is consistent with 
the idea that shear thickening
is caused by the development of the contact network~\citep{Seto_2013a}.


These results can be directly compared with those for the extensional flow simulation.
As seen in \figref{fig_microstructure}\,(\textit{b}),
even below thickening ($\dot{\varepsilon}/\dot{\varepsilon}_0 = 0.01$),
there is no long-range correlation in $g(\bm{r})$.
The distribution pattern has
horizontal and vertical mirror symmetries and no vorticity skews the correlation in the extensional flow.
The distribution pattern 
does not change much above thickening ($\dot{\varepsilon}/\dot{\varepsilon}_0 = 0.02$).
But a clear difference is present
in the angular contact distribution $g_{\mathrm{c}}(\theta)$~(\figref{fig_microstructure}\,(\textit{d})).
A flame-shaped distribution transforms into  
a fan-shaped distribution at the extensional thickening transition.
Thus, just below the transition,
we can find contacting particles only around the directions of the compression axis;
nevertheless,
the width of the flame shape indicates 
that, differently from what we observed under shear,
the contact chains do not correspond to stable ordered chains of particles.
Rather, they are constantly rebuilt among new neighbouring particles.
Such contact chains which are 
roughly parallel and oriented along the compression axis
do not contribute to the viscosity significantly~(\figref{fig_kappa_mono}).
By contrast, above the thickening transition,
contacting particles can be found in all directions,
even in the directions of the extension axis ($\theta = 0$ and $\pi$);
such distribution suggests an anisotropic network structure for the pattern of contacts,
which enhances the viscosity. 
Thus, we can describe the essence of extensional thickening 
as a \emph{contact-chain to contact-network} transition.

The fact that such a transition occurs in extensional flows
without significant change of the long-range correlation (always absent)
indicates that also in simple shear flows
the main responsible for thickening is
the contact-chain to contact-network transition.
Indeed, while we observe a concurrent order-disorder structural transition
in monodisperse suspensions under shear,
this is not present in strongly bidisperse suspensions,
which nevertheless display a strong thickening behaviour.

\section{Conclusions}

We numerically explored the non-Newtonian character of dense suspensions,
which has a different origin from that of viscoelastic fluids.
This character is manifested in three main aspects:
rate dependence, non-dissipative responses, and flow-type dependence.
Analysing thickening in both extensional and simple shear flows,
we were able to confirm that 
the contact-chain to contact-network transition is its main cause.
Non-dissipative responses, such as normal stress differences,
are present in any flow regime.
Flow-type dependence is evident in monodisperse suspensions
below thickening,
where ordering occurs under simple shear.
Preventing ordering through thickening or polydispersity hinders 
(but does not cancel)
the flow-type dependence.
%


\section*{Acknowledgements}

The authors acknowledge support from the Okinawa Institute of Science and Technology Graduate University.
The research of Ryohei Seto is partially supported by JSPS KAKENHI Grant Number JP17K05618.

\bibliographystyle{jfm}

\end{document}